\documentclass[aps,prd,letterpaper,preprint,preprintnumbers,nofootinbib,superscriptaddress]{revtex4-1}

\usepackage{amsmath,amssymb}
\usepackage{graphicx,multirow}
\usepackage{booktabs}
\usepackage{slashed}
\usepackage{units}
\usepackage[hyperfootnotes=false]{hyperref}
\usepackage[usenames, dvipsnames]{color}

\newcommand{\dd}{\ensuremath{\text{d}}}
\newcommand{\ii}{\ensuremath{\text{i}}}

\newcommand{\beq}{\begin{eqnarray}}% can be used as {equation} or  {eqnarray}
\newcommand{\eeq}{\end{eqnarray}}
\newcommand{\bea}{\begin{eqnarray}}% can be used as {equation} or  {eqnarray}
\newcommand{\eea}{\end{eqnarray}}

\begin{document}
%%%%%%%%%%%%%%%%%%%%%%%%%%%%%%%%%%

\preprint{UCI-TR-2021-29}

\title{Excited Q-Balls}

\author{Yahya Almumin}
\email{yalmumin@uci.edu}
\affiliation{Department of Physics and Astronomy, 
University of California, Irvine, CA 92697-4575, USA
}

\author{Julian Heeck}
\email{heeck@virginia.edu}
\affiliation{Department of Physics, University of Virginia,
Charlottesville, Virginia 22904-4714, USA}

\author{Arvind Rajaraman}
\email{arajaram@uci.edu}
\affiliation{Department of Physics and Astronomy, 
University of California, Irvine, CA 92697-4575, USA
}

\author{Christopher B.~Verhaaren}
\email{verhaaren@physics.byu.edu}
\affiliation{Department of Physics and Astronomy, Brigham Young University, Provo, UT, 84602, USA
}

\begin{abstract}
Complex scalars in $U(1)$-symmetric potentials can form stable Q-balls, non-topological solitons that correspond to spherical bound-state solutions. If the $U(1)$ charge of the Q-ball is large enough, it can support a tower of unstable radial excitations with increasing energy. Previous analyses of these radial excitations were confined to fixed parameters, leading to excited states with different charges $Q$. In this work, we provide the first characterization of the radial excitations of solitons for fixed charge, providing the physical spectrum for such objects. We also show how to approximately describe these excited states analytically and predict their global properties such as radius, energy, and charge. This enables a complete characterization of the radial spectrum. We also comment on the decay channels of these excited states. 
\end{abstract}
%%%%%%%%%%%%%%%%%%%%%%%%%%%%%%%%%
%%%%%%%%%%%%%%%%%%%%%%%%%%%%%%%%%

\maketitle

%%%%%%%%%%%

\section{Introduction}

Q-balls arise as spherically-symmetric localized classical solutions to the Lagrange equations of a complex scalar field $\phi(\vec{x},t)$ in a $U(1)$-symmetric potential that includes attractive interactions~\cite{Coleman:1985ki,Lee:1991ax}. While these solutions are intrinsically interesting examples of non-topological solitons, they may also be directly related to some of the big questions of what lies beyond the Standard Model of particle physics. They have been shown to arise in several motivated extensions of the standard model, such as supersymmetry~\cite{Kusenko:1997zq,Enqvist:1997si}, extra dimensions~\cite{Demir:2000gj,Abel:2015tca}, and hidden sectors with QCD-like confinement~\cite{Bishara:2017otb,Bishara:2021fag}. They can produce intriguing macroscopic dark matter candidates~\cite{Kusenko:1997si,Kusenko:1997vp,Kusenko:2001vu,Ponton:2019hux,Bai:2019ogh,Bai:2021mzu}, may be related to the observed asymmetry between matter and antimatter~\cite{Enqvist:1997si,Krylov:2013qe}, and can give rise to observable gravitational waves~\cite{Kusenko:2008zm,Croon:2019rqu}. 

The Q-ball ground-state configuration has the lowest energy for a given $U(1)$ charge, rendering it stable~\cite{Coleman:1985ki}.
Exact analytical solutions can only be obtained for special\textemdash typically unphysical\textemdash potentials, other cases have to be approached numerically or via analytic approximations. The latter case is particularly fruitful for large Q-balls, with the large radius acting as a good expansion parameter. In Ref.~\cite{Heeck:2020bau}, excellent analytic approximations for Q-balls in a sextic potential were recently provided, which essentially eliminate the need to solve the differential equations numerically. This paves the way for further phenomenological studies to include Q-balls states accurately without continually solving the nonlinear systems that define them.

Spherically symmetric ground-state configurations of the form $\phi (\vec{x},t) = \tilde\phi( |\vec{x}|) e^{\ii \omega t}$~\cite{Coleman:1985ki} are typically the simplest solutions and those of most interest for phenomenological studies, given their stability. Nevertheless, the study of unstable excited states is useful and necessary because Q-ball formation and scattering will generally also produce excited states that eventually relax into the ground state~\cite{Battye:2000qj,Multamaki:2001az,Multamaki:2002hv,Kusenko:2008zm,Tsumagari:2009na,Hiramatsu:2010dx}. The details of the excitation spectrum are therefore important, e.g.~for Q-ball dark matter, just like the study of excited nuclei and atoms.
In the early universe, the scalars $\phi$ and Q-balls are expected to be in thermal equilibrium~\cite{Griest:1989cb,Griest:1989bq,Postma:2001ea}; excited states will then be in equilibrium as well, with their abundance suppressed by their larger mass. It is therefore crucial to know the mass gap or spectrum of Q-balls of fixed charge $Q$ to ascertain their relevance in the cosmological evolution.
Similarly, any scattering of or off Q-balls generically produces excited states unless kinematically forbidden due to a large mass gap.
Furthermore, the lifetimes of excited states could be long if the available phase space is suppressed: on general grounds, the lifetime  of an excited state will scale inversely with some power of the energy gap $\Delta E$ to the ground state due to the reduced phase space; for a compressed excitation spectrum, the lifetimes of the excited states could hence be large enough to be of phenomenological interest~\cite{Multamaki:2001az}. 

Building on the ground-state work of Ref.~\cite{Heeck:2020bau}, we therefore study excited Q-ball states, providing useful analytical approximations to the exact solutions, as shown by our comparison to the numerical data.
Excited Q-balls with angular momentum $J\neq 0$, have been discussed in Refs.~\cite{Volkov:2002aj,Kleihaus:2005me} and shown to have energies that exceed the non-rotating ground-state energy by at least $\sim 20\%$ for the same charge $Q$. These initial studies give a reasonable idea of the Q-ball spectrum in terms of angular momentum, albeit only for a small region in parameter space.
In addition to these rotational excitations, Q-balls exhibit spherically symmetric \emph{radial} excitations with vanishing angular momentum, discovered in Refs.~\cite{Friedberg:1976me,Volkov:2002aj}.
Numerical solutions for the first 23 radial excitations for a fixed frequency $\omega$  have been computed and discussed in Ref.~\cite{Mai:2012cx}.\footnote{The excited states of \emph{gauged}  Q-balls for the sextic potential are explored in Ref.~\cite{Loginov:2020xoj}.}  Because $\omega$ was fixed, all these excitations have different charges $Q$, so they cannot be interpreted as physical excitations of each other or a specific ground state soliton. In fact, to the best of our knowledge, no study of the radial excitation spectrum for a fixed $Q$ exists in the literature. It has, therefore, not yet been established whether the lowest-lying Q-ball excitation is radial or rotational.
Building upon and generalizing aspects of previous work, our efforts below are able to characterize all radial excited states for the complete family of sextic potentials and all phases, providing the full radial spectrum. Despite our analysis being performed for one particular scalar potential, we expect our results to be valuable more generally and adaptable to other potentials.

In the following section we review the theory of ground state Q-balls to establish notation, conventions, and the definitions of physical characteristics like charge and energy. We also review the different types of stability these solitons may exhibit.  Section~\ref{sec:excited} motivates an analytic form for excited state Q-balls and methods to fully characterize them in terms of the potential parameters. Approximations to the physical characteristics of the excited states are obtained in Sec.~\ref{sec:discussion} as well as a discussion of their stability and possible decay channels. In Sec.~\ref{sec:GenPotential} we outline how our results are expected to generalize to other potentials. We conclude in Sec.~\ref{s.con}.

%%%%%%%%%%%%%%%%%%%%
%%%%%%%%%%%%%%%%%%%%
\section{Review of Ground-State Q-Balls}
\label{sec:ground}

We recount the well-known case of ground-state Q-balls to establish our notation, following closely Ref.~\cite{Heeck:2020bau}. The Lagrangian for the complex scalar $\phi$ is simply
\begin{equation}
\mathcal{L}=\left|\partial_\mu\phi \right|^2-U(|\phi|),
\end{equation}
with a $U(1)$ invariant potential $U$ that has to fulfill a number of conditions to support Q-ball solutions. Since we do not want to break the $U(1)$ symmetry, we require $\langle\phi\rangle=0$ in the vacuum; we normalize the potential energy to zero in the vacuum by setting $U(0)=0$ and enforce that the vacuum is a stable minimum of the potential by demanding
\begin{equation}
\left.\frac{\dd U}{\dd |\phi|}\right|_{\phi=0}=0\,, \ \ \ \ \left.\frac{\dd^2 U}{\dd \phi\, \dd\phi^\ast}\right|_{\phi=0}=m_\phi^2 > 0\,,
\end{equation}
where $m_\phi$ is the mass of the complex scalar. Coleman~\cite{Coleman:1985ki} showed that nontopological solitons, Q-balls, exist when the function $U(|\phi|)/|\phi|^2$ has a minimum at $|\phi|=\phi_0/\sqrt{2}>0$\footnote{One can in fact consider Q-balls with only a thick-wall limit~\cite{Kusenko:1997ad,PaccettiCorreia:2001wtt} in certain potentials. In this case $\phi_0$ is not defined.} such that 
\begin{equation}
0\leq\sqrt{\frac{2U(\phi_0/\sqrt{2})}{\phi_0^2}}\equiv \omega_0<m_\phi\,.\label{e.Omega0}
\end{equation}
In this case, the spherically symmetric Q-ball solutions  take the form
\begin{equation}
\phi (x^\mu)=\frac{\phi_0}{\sqrt{2}}f(r) e^{\ii\omega t}\,,\label{e.fdef}
\end{equation}
where $f(r)$ is a dimensionless function of the radius $r$ and $\omega$ is a  constant internal frequency that is restricted to the region $\omega\in (\omega_0, m_\phi)$.
Switching to a  dimensionless radial coordinate $\rho$ defined by
\beq
\rho=r\sqrt{m_\phi^2-\omega_0^2}\,,
\eeq
we end up with the effective Lagrangian for $f(\rho)$
\beq
L=\frac{4\pi\phi_0^2}{\sqrt{m_\phi^2-\omega_0^2}}\int \dd\rho\; \rho^2\left[-\frac12f^{\prime2}+V(f) \right] ,
\eeq
where a prime denotes a derivative with respect to $\rho$ and the effective potential $V(f)$ is defined via
\begin{align}
V(f) = \frac{1}{m_\phi^2 - \omega_0^2} \, \left( \frac{\omega^2}{2} f^2 - \frac{U(f\phi_0/\sqrt{2})}{\phi_0^2}\right) .
\end{align}
The Euler-Lagrange equation,
\begin{align}
f'' +\frac{2}{\rho} f' + \frac{\dd V }{\dd f} = 0 \,,
\label{eq:differential_equation}
\end{align}
is then equivalent  to a particle of position $f$ moving in the potential $V(f)$ while subject to friction, $\rho$ corresponding to the time coordinate in this analogy~\cite{Coleman:1985ki}. The boundary conditions are $f'(0)=0$ and $f(\rho\to\infty) = 0$ in order to obtain localized solutions.
The charge $Q$ and energy $E$ of the Q-ball are given by the integrals
\begin{align}
\begin{split}
Q &=\frac{4\pi \omega \phi_0^2}{(m_\phi^2 - \omega_0^2)^{3/2}} \int \dd\rho\,\rho^2f^2 \,,\\
E&=\omega Q + \frac{4\pi\phi_0^2}{3\sqrt{m_\phi^2 - \omega_0^2}}\int \dd\rho\,\rho^2f^{\prime2} \,,
\end{split}
\label{eq:QandE}
\end{align}
which satisfy the differential equation $\dd E/\dd \omega = \omega \dd Q/\dd \omega$~\cite{Friedberg:1976me,Lee:1991ax,Heeck:2020bau}.

\begin{figure}[t]
\includegraphics[width=0.75\textwidth]{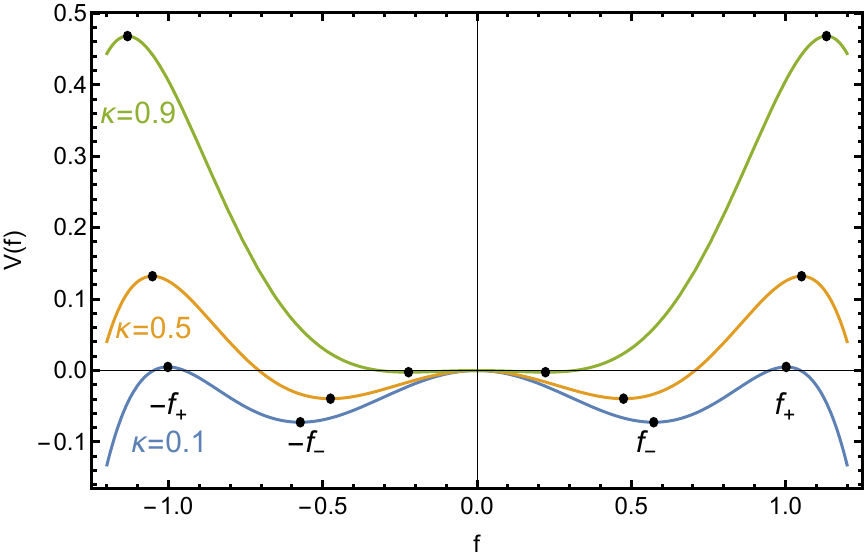}
\caption{
Potential $V(f)$ for $\kappa=0.1$ (blue), $\kappa=0.5$ (tan), and $\kappa=0.9$ (green). The extrema  $(\pm) f_{\pm}$ are shown as black dots.}
\label{fig:EffPot}
\end{figure}

To be more concrete, let us consider the sextic potential 
\begin{align}
U(\phi) = m_\phi^2 |\phi|^2 - \beta |\phi|^4+\frac{\xi}{m_\phi^2} |\phi|^6 \,,\label{e.sextic}
\end{align}
which is non-renormalizable but can easily be UV-completed by introducing heavier particles. 
Replacing the parameters $\beta$ and $\xi$ by $\phi_0$ and $\omega_0$, this potential leads to the function $V$
\begin{align}
V(f) = \frac{1}{2} f^2 \left[ \kappa^2 - (1-f^2)^2\right], \quad \text{ with } \quad \kappa^2 \equiv  \frac{\omega^2 - \omega_0^2}{m_\phi^2 - \omega_0^2} \,,
\end{align}
where $\kappa\in (0,1)$~\cite{Heeck:2020bau}. The potential has extrema at $f=0$ and 
\beq
f^2_\pm=\frac13\left(2\pm\sqrt{1+3\kappa^2} \right) .\label{e.fpm}
\eeq
 As shown in Fig.~\ref{fig:EffPot}, the $\pm f_+$ are maxima while $\pm f_-$ are local minima. 

The differential equation~\eqref{eq:differential_equation} depends only on the parameter $\kappa$, which must therefore determine the radius of the Q-ball.
For ground-state solutions we demand $f(\rho)$ to be a monotonic function and define the radius $R^*$ through $f''(\rho=R^*) = 0$. 
In the limit of large $R^*$ or small $\kappa$, the friction term $2f'/\rho$ in Eq.~\eqref{eq:differential_equation} becomes negligible, allowing us to find the approximate profile $f(\rho)$ in the form of the so-called transition function
\begin{align}
f_T (\rho ) = \frac{1}{\sqrt{1+2 e^{2(\rho-R^*)}}} \,,
\label{eq:fT}
\end{align}
which becomes an exact solution in the limit $R^*\to \infty$~\cite{Heeck:2020bau}. 
For small $\kappa$, this transition from $f=f_+\simeq 1$ to $f=0$ reduces the energy by $1/(2R^\ast)$ due to friction. The total energy loss must be the difference between the initial maximum, with energy given at leading order in $\kappa$ by 
\begin{align}
V(f_+)=\frac{2+\sqrt{1+3\kappa^2}}{27}\left(\sqrt{1+3\kappa^2}-1+3\kappa^2 \right)=\frac{\kappa^2}{2}+\mathcal{O}(\kappa^4)\,,
\label{eq:Vplus}
\end{align}
 and the final maximum at $V(0)=0$. This implies that to leading order in small~$\kappa$:
\beq
\frac{1}{2R^\ast}=\frac{\kappa^2}{2}\quad\Rightarrow\quad R^\ast=\frac{1}{\kappa^2}\,.
\label{eq:R_vs_kappa_ground}
\eeq
This relation between radius and small $\kappa$ is a good approximation for all stable ground-state solutions~\cite{Heeck:2020bau}.
With these predictions for the radius and profile we can also obtain analytic expression for energy and charge via Eq.~\eqref{eq:QandE}.

\subsection{Ground State Stability}
\label{sec:ground_state_stability}

Discussions of stability regarding Q-ball ground states are often divided into three categories: absolute stability, classical stability, and stability to fission~\cite{Tsumagari:2008bv}. The absolute stability criterion is the simplest: it is the requirement that the soliton solution of charge $Q$ have a lower energy than $Q$ free scalar particles. In other words, the Q-ball energy $E$ must satisfy $E<m_\phi Q$. This is referred to as absolute stability because it is stable against both classical and quantum effects~\cite{Lee:1991ax}. 
In our sextic potential, ground-state solutions are absolutely stable for $\kappa\lesssim 0.84$~\cite{Heeck:2020bau}, so in particular in the thin-wall regime $\kappa \ll 1$.

Classical stability is taken to mean that the soliton is stable against perturbations; if  perturbations can grow without bound, then the solution is said to be unstable. It has been shown that when 
\beq
\frac{\omega}{Q}\frac{\dd Q}{\dd\omega}\leq0\,,
\eeq
the corresponding Q-ball solutions are classically stable~\cite{Friedberg:1976me,Lee:1991ax}. We are free to choose both $\omega>0$ and $Q>0$, so the more significant condition for that choice is
\beq
\frac{\dd Q}{\dd\omega}\leq0\,. \label{e.Qstable}
\eeq
In the sextic potential, this requirement is automatically satisfied in the absolutely-stable region with $\kappa\lesssim 0.84$ and hence  a weaker criterion.
 It is worth noting that the derivation of Eq.~\eqref{e.Qstable}  also demonstrates that the radial profile for these classically stable ground-state solitons have no nodes~\cite{Friedberg:1976me,Coleman:1985ki,Lee:1991ax}. 

The final stability criterion is related to a soliton breaking up into smaller solitons and free particles, something like the fission of a nucleus. It was argued in~\cite{Lee:1991ax} that the condition in \eqref{e.Qstable} also prevents Q-ball fission. This can be seen from the following argument given in~\cite{Gulamov:2013cra}. Using $\dd E/\dd Q=\omega$, we can write
\beq
\frac{\dd\omega}{\dd Q}=\frac{\dd}{\dd Q}\frac{\dd E}{\dd Q}=\frac{\dd^2E}{\dd Q^2}~.
\eeq
This means that when $\frac{\dd Q}{\dd\omega}<0$ it follows that
\beq
\frac{\dd^2E}{\dd Q^2}<0~.\label{eq:FissionInEq}
\eeq

Consider the possible fission of a Q-ball of charge $Q_1+Q_2$ that breaks into two solitons of charge $Q_1$ and $Q_2$, respectively. We integrate the inequality in \eqref{eq:FissionInEq} from some intermediate charge $Q_i$ up to $Q_i+Q_2$:
\beq
\int_{Q_i}^{Q_i+Q_2}\dd Q\, \frac{\dd^2E}{\dd Q^2}<0\;\Rightarrow\; \left.\frac{\dd E}{\dd Q}\right|_{Q_i+Q_2}< \left.\frac{\dd E}{\dd Q}\right|_{Q_i}~.
\eeq
We then integrate this resulting inequality with respect to the intermediate charge $Q_i$ from 0 to $Q_1$. Because $E(0)=0$, this leads to 
\beq
E(Q_1+Q_2)<E(Q_2)+E(Q_1)~,
\eeq
which means that fission is energetically forbidden if Eq.~\eqref{e.Qstable} is satisfied.

%%%%%%%%%%%%%%%%%%%%%%%%%%%%
%%%%%%%%%%%%%%%%%%%%%%%%%%%%
\section{Excited State Solutions}
\label{sec:excited}

Heretofore we have assumed that the rolling particle always has $f\geq0$. However, one can imagine the particle beginning from rest and then rolling down the hill in positive $f$, over the maximum at $f=0$, up the hill in negative $f$, and back toward rest at $f=0$. Such a trajectory satisfies the Q-ball boundary conditions and leads to a localized solution. The corresponding solitons are said to belong to the first excited state of some Q-ball~\cite{Volkov:2002aj,Mai:2012cx}, for reasons which are made clear below. 
More generally, the particle can roll back and forth several times before settling down at $f=0$, which defines an entire tower of excited states. We label the excited states by the number $N$ of transitions through $f=0$, with $N=0$ being the $f\geq0$ ground state.

The only way to enable the particle to roll a longer distance is for it to begin with larger potential energy. Thus, for a fixed $\kappa$, energy conservation suggests that the particle's initial position $f(0)$ moves closer to the maximum $f_+$ for increasing $N$. Starting closer to the maximum implies the particle's velocity is nearly zero for a longer time during which the friction term decreases. When larger scale motion eventually begins, the friction has deceased sufficiently for the particle to complete the larger number of back-and-forth transitions before coming to rest. In short, by delaying the onset of motion in transition toward $f=0$ the particle loses less energy to friction. Qualitatively, this shows that there is an infinite tower of excited states for a given $\kappa$ with each higher $N$ trajectory beginning slightly closer to the maximum $f_+$. It also suggests that the radius of the Q-ball (the `time' of the first transition) grows with $N$ for large enough $N$, which is confirmed by the numerical results that follow.

Notice that for a fixed $\kappa$, the $N=0$ and $N>0$ states do not have the same $Q$~\cite{Mai:2012cx}, so they should not be regarded as excitations of each other. This follows from the rolling particle analogy upon realizing that the $N>0$ states necessarily start off closer to $f_+$ in order to have sufficient energy for their additional transitions, which results in a larger integral $\int \dd\rho\,\rho^2 f^2$ and thus a larger $Q$ due to Eq.~\eqref{eq:QandE}.
In addition, stable Q-balls require $\dd Q/\dd \omega < 0$~\cite{Tsumagari:2008bv}, or $\dd Q/\dd \kappa <0$ in our notation, so $Q$ \emph{decreases} with increasing $\kappa$.
Therefore, to have the $N=1$ soliton's $Q$ match a $N=0$ soliton $Q$, the $\kappa$ of the excited state must be \emph{larger} than the ground state $\kappa$. Generally, exciting a Q-ball from $N$ to $N+1$ requires an increase in $\kappa$ or $\omega$ to keep $Q$ the same.
The energy (Eq.~\eqref{eq:QandE}) of the excited state for a fixed $Q$ then unavoidably increases, both due to the larger $\omega$ and because the surface integral  $\int \dd\rho\,\rho^2 (f')^2$ increases with every back-and-forth transition, justifying the terminology that each larger $N$ soliton is an excited state of an $N=0$ Q-ball.

\subsection{Profiles}
\label{sec:profiles}

\begin{figure}[t]
\includegraphics[width=0.49\textwidth]{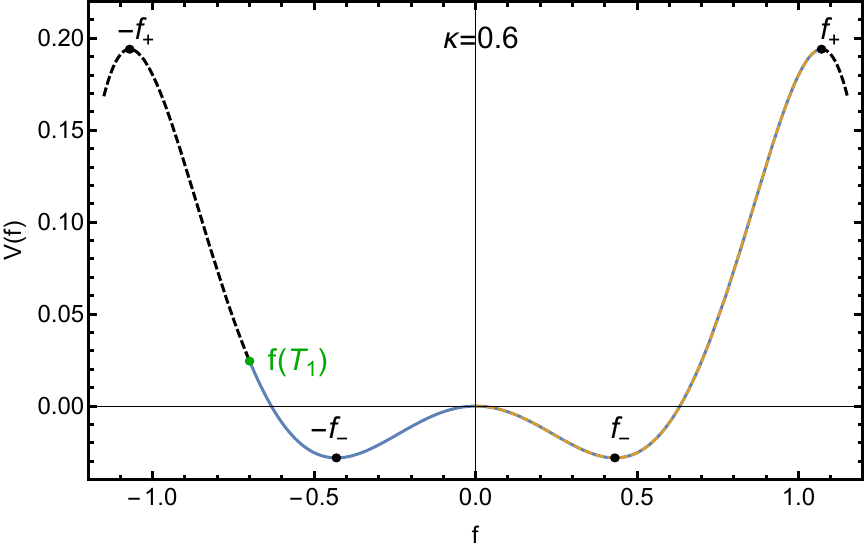}
\includegraphics[width=0.49\textwidth]{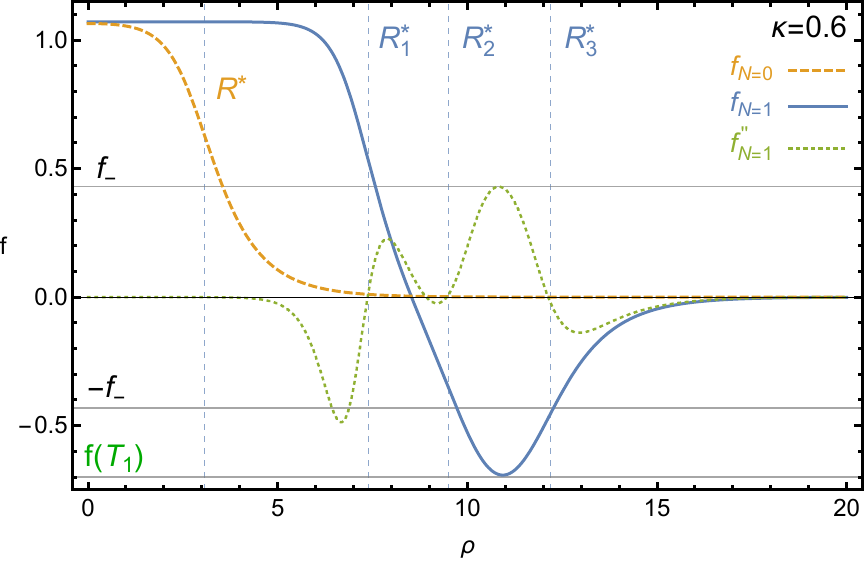}
\caption{ Effective potential $V(f)$ (left) and $N=1$ excited state profile $f(\rho)$ (right). The solid line in the potential denotes the path along which the particle rolls. The locations of the relevant extrema of $V(f)$, $f_\pm$ are denoted in each plot, as is the location of the turning point $f(T_1)$.
 }
\label{fig:N1EffPot}
\end{figure}

Most of the characteristics of excited-state trajectories may be understood by considering the $N=1$ example shown in Fig.~\ref{fig:N1EffPot}. Even for this $\kappa=0.6$ example (which is not close to the $\kappa\to0$ thin-wall limit) the particle begins at rest near the local maximum at $f_+$. After remaining for some time near the maximum, it transitions quickly down the potential hill. In fact, it is more useful to consider the particle as having approximately three transitions. The first is similar to the ground state Q-ball (the dashed tan line in the figure), rolling from $f_+$ to $f=0$. Instead of stopping, however, it then transitions uphill in negative $f$. The particle has lost energy due to friction along its way, so the turning point $f(T_1)$ is well below the maximum at $-f_+$. The particle finally transitions back down from the turning point and comes to rest at on the maximum at $f=0$. The profile in physical space that corresponds to this particle trajectory is shown as the solid blue line in the right panel of Fig.~\ref{fig:N1EffPot}.

The $N=0$ (dashed tan) and $N=1$ (solid blue) profiles on the right side of Fig.~\ref{fig:N1EffPot} illustrate several points. First, we see that the initial value $f(0)$ of the excited state profile is larger than the $f(0)$ for the ground state, as expected. Consequently, the excited state remains near the $f_+$ maximum until a larger $\rho$, allowing it to lose much less energy as it rolls. For the ground state there is only one point with $f''=0$, which defines the radius $R^\ast$. However, for the excited state there are four points where $f''=0$ (small dashed green). Three of these are near to where $f\approx \pm f_-$, and we have labeled these $R_1^\ast$, $R_2^\ast$, and $R^\ast_3$. The remaining zero of $f''$ is near to $f\approx0$. One can see that the $f''$ roots associated with $f=0$ and $R^\ast_2$ are somewhat marginal, and indeed as $\kappa$ increases, these two roots no longer have real solutions. 

\begin{figure}[t]
\includegraphics[width=0.49\textwidth]{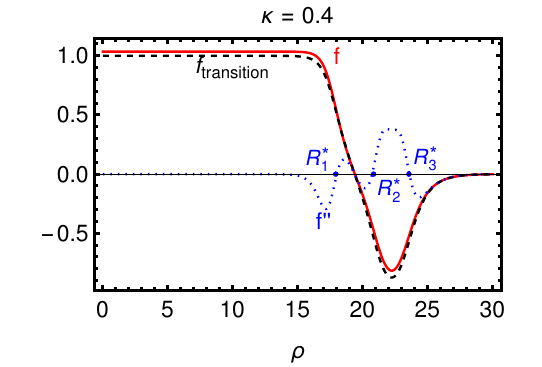}
\includegraphics[width=0.49\textwidth]{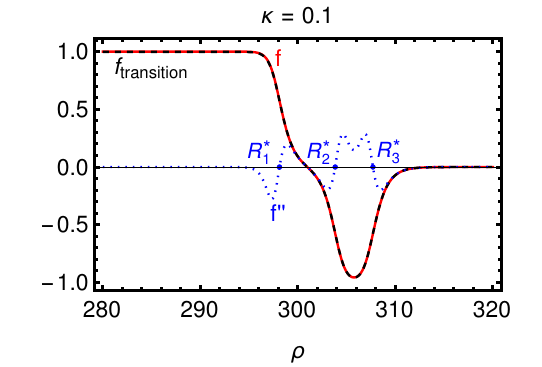}
\caption{
Numerical profiles $f(\rho)$ in red for $\kappa = 0.4$ (left) and $\kappa =0.1$ (right).
Also shown in dashed black is the analytic approximation $f_\text{transition}$ from Eq.~\eqref{eq:fT1}; the latter  depends on the three radii $R^\ast_j$ defined by $f''(R^\ast_j)=0$ (not the one at $f=0$). The agreement between numerical and transition profile becomes better for small $\kappa$.}
\label{fig:N1profiles}
\end{figure}

Describing the motion as three transitions\textemdash each with a defining radius at which $f''=0$ but $f\neq0$\textemdash allows us to make an ansatz for $f$ as the product of transition functions of the form given in Eq.~\eqref{eq:fT}:
\beq
f_{N=1}= \left[f_T(\rho,R^\ast_1)-f_T(-\rho,-R^\ast_2) \right]f_T(\rho,R^\ast_3)\,,
\label{eq:fT1}
\eeq
where $R_1^\ast<R_2^\ast<R_3^\ast$. Just like the transition profile of the ground state, this is only expected to be a reasonable form for large radii or small $\kappa$.
In Fig.~\ref{fig:N1profiles}, we show the exact numerical profiles $f_{N=1}$ for $\kappa=0.4$ and $0.1$ together with the above transition-function ansatz, fitting the three radii to the numerical data. For small $\kappa$, the agreement is remarkable and illustrates that the $N=1$ profile can be completely specified by the three radii $R^\ast_{1,2,3}$. At larger $\kappa$, this ansatz becomes increasingly inadequate\textemdash just like in the ground-state case.
We return to the large $\kappa$ regime in subsection~\ref{sec:thick}, for now we focus on small $\kappa$.

The extension of Eq.~\eqref{eq:fT1} for the $N$th excited state is simply
\begin{align}
f_N= \left[f_T(\rho,R^\ast_{N,1})-f_T(-\rho,-R^\ast_{N,2}) \right]\cdots\left[f_T(\rho,R^\ast_{N,2N-1})-f_T(-\rho,-R^\ast_{N,2N})\right]f_T(\rho,R^\ast_{N,2N+1}) \,,
\label{e.Nansatz}
\end{align}
specified by $2 N + 1$ radii $R^\ast_{N,n}$, $n=1,2,3,\ldots, 2 N +1$. The agreement with the numerical profiles is similar to Fig.~\eqref{fig:N1profiles} and generally excellent for small $\kappa$.
Though the numerical results necessarily assume a particular potential, the form of the excited state profile in terms of transition functions can be applied to any potential.

\subsection{Thin-Wall Limit}
\label{sec:thin}

The product of transition functions in Eq.~\eqref{e.Nansatz} is an empirically successful approximation for the excited-state profiles at small $\kappa$. It requires $2 N + 1$ radii $R^\ast_{N,n}$, $n=1,2,3,\ldots, 2 N +1$, as input, which can only depend on the small parameter $\kappa$ and necessarily diverge in the small $\kappa$ limit just like in the ground-state case. 
In this section, we derive the relations $R^\ast_{N,n}(\kappa)$ for small $\kappa$.

For the ground state, a small $\kappa$ transition reduced the particle energy by $1/(2R^\ast)$~\cite{Heeck:2020bau}. The first excited state with its three transitions is then expected to 
lose an energy of approximately ${3/( 2R^\ast)}$ due to friction. The total energy loss must be the difference between the initial maximum, with energy given at leading order in $\kappa$ by
Eq.~\eqref{eq:Vplus}, and final maximum at $V(0)=0$. This implies that to leading order in small $\kappa$
\beq
\frac{3}{2R^\ast}=\frac{\kappa^2}{2}\;\Rightarrow\; R^\ast=\frac{3}{\kappa^2}\,.
\eeq
We therefore expect all three radii of the first excited state, $R^\ast_{1,n}$, to be around $3/\kappa^2$, with only small differences $\Delta_n\equiv R^\ast_{1,n}-R^\ast_{1,n-1}$ that are subleading in small $\kappa$.
Similarly, the general $N$th excited state is expected to have radii  that are clustered around
\begin{align}
R^\ast_{N,n}=\frac{2N+1}{\kappa^2} \,,
\label{eq:leading_radius}
\end{align}
further split up by terms that are subleading in small $\kappa$, to be discussed next.

To estimate the $2N+1$ individual radii beyond Eq.~\eqref{eq:leading_radius} we need to determining the radial distance between them. In the thin-wall limit we may neglect friction. Then, the ``time" it takes for a particle to roll from one point on the effective potential to another is given by
\beq
\rho_2-\rho_1=\int_{f_1}^{f_2}\frac{\dd f}{\sqrt{2\mathcal{E}-2V(f)}}\,,
\label{e.IntApprox}
\eeq
where $f(\rho_i)=f_i$ and
\beq
\mathcal{E}=\frac12f^{\prime2}+V(f)\,,
\eeq
is the conserved energy of the particle~\cite{Heeck:2020bau}. In this same frictionless limit the particle's equation of motion is
\beq
f''=-\frac{\dd V}{\dd f}\,.
\eeq
This means the values of $f$ for which $f''=0$ (which define the radii) are exactly the extrema of $V(f)$. More particularly, as is clear from Fig.~\ref{fig:N1EffPot}, we are interested only in the extrema that occur when $f\neq0$. Then, to find the distance between the radii, we consider two types of trajectories. Those that pass through $f=0$, such as from $f=f_-$ to $f=-f_-$, and those that run from the potential minimum to a turning point, such as from $f=-f_-$ up the potential hill to the turning point $f(T_i)$ and then back to $f=-f_-$. 

In the thin-wall (or small $\kappa$) limit, only a small range of $f\sim f_+$ correspond to $V(f)>0$. As the particle loses energy to friction and must end at $V(0)=0$, the particle's initial location must be in this small region $f\lesssim f_+$. This also ensures that it does not begin to transition until the friction is largely absent so that the energy loss is small. Thus, the value of $\mathcal{E}$ is also small, with $f$ beginning on the potential with $f'=0$. The be more precise,
\beq
\mathcal{E}\approx V(f_+)=\frac{\kappa^2}{2}+\mathcal{O}(\kappa^4)\,.
\eeq
This means that we can expand the ``time" integral as
\begin{align}
&\int \dd f\left[\frac{1}{f(1-f^2)}+\frac{f^2\kappa^2-2\mathcal{E}}{2f^3(1-f^2)^3}+\mathcal{O}(\kappa^4) \right]\nonumber\\
&=\frac12\left(1+\frac12\kappa^2+3\mathcal{E} \right)\ln\frac{f^2}{1-f^2}+\frac{3-2f^2}{8(1-f^2)^2}\kappa^2-\frac{2-9f^2+6f^4}{4f^2(1-f^2)^2}\mathcal{E}+\mathcal{O}(\kappa^4)\,.\label{e.timeExpansion}
\end{align}
It remains to specify the limits of integration. We first consider the trajectories that change the sign of $f$. The relevant interval is $-f_-$ to $f_-$, which provides estimates for $R^\ast_{N,\text{even}}-R^\ast_{N,\text{odd}}$. However, as Eq.~\eqref{e.timeExpansion} is a function of $f^2$ and the limits of integration are even, it seems that the integral vanishes. The subtlety is the singularity at $f=0$. 

Our approximation is not valid when $f\approx 0$. That is, when $f^2\sim\kappa^2$ the expansion in Eq.~\eqref{e.timeExpansion} breaks down. But, in this regime the potential can be well approximated by the quadratic term
\beq
V(f)\approx-\frac12(1-\kappa^2)f^2+\mathcal{O}(f^4)\,.
\eeq
 We can then integrate
 \beq
 \int \frac{\dd f}{\sqrt{2\mathcal{E}+(1-\kappa^2)f^2}}=\frac{1}{\sqrt{1-\kappa^2}}\ln\left[ f(1-\kappa^2)+\sqrt{1-\kappa^2}\sqrt{2\mathcal{E}+f^2(1-\kappa^2)}\right] .
 \eeq
 We assume that these two regimes are joined at some value $f=f_j$. We need $f_j>\kappa$, so that the expansion for $f>f_j$ is justified. A consistent choice is $f_j=\sqrt{\kappa}$. Then we find
\begin{align}
 \int_0^{f_j} \frac{\dd f}{\sqrt{2\mathcal{E}+(1-\kappa^2)f^2}}&=\frac{1}{\sqrt{1-\kappa^2}}\ln\frac{ f_j\sqrt{1-\kappa^2}+\sqrt{2\mathcal{E}+f_j^2(1-\kappa^2)}}{\sqrt{2\mathcal{E}}}\nonumber\\
 &=\frac12\ln\frac{2f_j^2}{\mathcal{E}}+\frac{\mathcal{E}}{2f_j^2}+\mathcal{O}(\kappa^2)\,.
 \end{align}
 The other part of the integration is from $f_j$ to $f_-$ and is found to be
 \begin{align}
&\int_{f_j}^{f_-} \dd f\left[\frac{1}{f(1-f^2)}+\frac{f^2\kappa^2-2\mathcal{E}}{2f^3(1-f^2)^3}+\mathcal{O}(\kappa^4) \right] =\frac12\ln\frac{f_-^2(1-f_j^2)}{f_j^2(1-f_-^2)}-\frac{\mathcal{E}}{2f_j^2}+\mathcal{O}(\kappa^2)\,.
\end{align}
Note that when these terms are combined that the term linear in $\mathcal{E}$ (and also in $\kappa$) cancels. The leading result is
\beq
\int_0^{f_-}\frac{\dd f}{\sqrt{2\mathcal{E}-2V(f)}}=\frac12\ln\frac{2f_-^2}{\mathcal{E}(1-f_-^2)}+\mathcal{O}(\kappa^2)\,.
\eeq
By multiplying by two, we obtain the full trajectory from $-f_-$ to $f_-$ and so
\beq
\Delta_\text{even}=\ln\frac{2f_-^2}{\mathcal{E}(1-f_-^2)}\,,
\eeq
is our leading order estimate of $R^\ast_{N,\text{even}}-R^\ast_{N,\text{odd}}$.

Before this can be used to estimate excited state radii we need to determine $\mathcal{E}$. The initial energy of the particle rolling the potential is taken to be approximately $V(f_+)$. The particle loses energy until is stops at $V(0)=0$. We estimate the energy loss by dividing the total energy equally among all the transitions, which for the $N$th excited state is $2N+1$ transitions. 
The energy at the beginning of the $n$th transition is given by
\beq
\mathcal{E}_n=V(f_+)\frac{2N+2-n}{2N+1}\approx\frac{\kappa^2}{2}\frac{2N+2-n}{2N+1}\,,
\eeq
where in the last expression we use the leading order in $\kappa$ result.

Finally, to determine $\Delta_\text{odd}$ we must estimate the ``time" it takes the particle to go from the minimum of the potential up to a turning point and back. This is equivalent to twice the integral from $-f_-$ up to the turning point $f_{T}$ of the particle. This turning point is determined by the equation
\beq
V(f_{T})=\frac{f_{T}^2}{2}\left[\kappa^2-\left(1-f_{T}^2\right)^2 \right]=\mathcal{E}_n\,.\label{e.turningPot}
\eeq
 Because of the symmetry of the potential, we consider the equivalent trajectory with $f>0$. In the $\kappa\to0$ limit $f_{T}\lesssim1$, so the $1-f^2$ terms in the denominator of Eq.~\eqref{e.timeExpansion} can invalidate the expansion. Since $f_{T}\to1$ as $\kappa\to0$ we parameterize the turning point location as
 \beq
 f_{T}=1-T\kappa \,.
 \eeq
 To leading order in $\kappa$ Eq.~\eqref{e.turningPot} becomes
 \beq
 \frac12(1-T^2)\kappa^2=\frac{\kappa^2}{2}\frac{2N+2-n}{2N+1}\,,
 \eeq
 which implies that
 \beq
 T=\sqrt{\frac{n-1}{2N+1}}\,.
 \eeq
Note here that while $n$ labels the transition, only odd $n$'s larger than one are actually associated with turning points. So, we label the $m$th turning point, with $n=2m+1$, as
\beq
T_m=\sqrt{\frac{2m}{2N+1}}\,.
\eeq
  We now need to consider the integral
\beq
\int_{f_j}^{f_{T_n}}\frac{\dd f}{\sqrt{2\mathcal{E}-2V(f)}}\,,
\eeq
in the $f\sim1$ limit. In this case we take $f_j=1-\kappa$ to ensure the expansion is justified.  We transform to the variable $h=1-f$ and find
\beq
\int^{\kappa}_{\kappa T}\frac{\dd h}{\sqrt{2\mathcal{E}-2V(h)}}\,,
\eeq
where
\beq
V(h)=\kappa^2\left(\frac12-h \right)-\frac12\left(4-\kappa^2 \right)h^2+\mathcal{O}(h^3)\,.
\eeq
By keeping only terms up to $h^2$ we account for all the $\kappa$ dependence and the dominant effects of $h$ in the $h\sim0$ limit. This integral is evaluated to be
\beq
\frac12\ln\frac{2\kappa+\sqrt{3\kappa^2+2\mathcal{E}}}{2\kappa T+\sqrt{(4T^2-1)\kappa^2+2\mathcal{E}}}+\mathcal{O}(\kappa)\,.
\eeq

The other part of the integration is from $f_-$ to $f_j$ and is found to be
 \begin{align}
&\int^{f_j}_{f_-} \dd f\left[\frac{1}{f(1-f^2)}+\frac{f^2\kappa^2-2\mathcal{E}}{2f^3(1-f^2)^3}+\mathcal{O}(\kappa^4) \right] =\frac12\ln\frac{1-f_-^2}{2\kappa f_-^2}+\frac{1-2\mathcal{E}/\kappa^2}{32}+\mathcal{O}(\kappa)\,.
\end{align}
Then twice the total integral gives the estimate of $R^\ast_{N,\text{odd}}-R^\ast_{N,\text{even}}$:
\beq
\Delta_\text{odd}=\frac{1-2\mathcal{E}/\kappa^2}{16}+\ln\left[\frac{1-f_-^2}{2\kappa f_-^2}\frac{2\kappa+\sqrt{3\kappa^2+2\mathcal{E}}}{2\kappa T+\sqrt{(4T^2-1)\kappa^2+2\mathcal{E}}}\right] .
\eeq

In summary, the radial distances $\Delta_n = R^\ast_{N,n}-R^\ast_{N,n-1}$ between radii are estimated to be
\begin{align}
\Delta_n&=
\begin{cases}
\ln\frac{2f_-^2}{\mathcal{E}_n(1-f_-^2)} & n\text{ even}\,,\\
\frac{1-2\mathcal{E}_n/\kappa^2}{16}+\ln\left[\frac{1-f_-^2}{2\kappa f_-^2}\frac{2\kappa+\sqrt{3\kappa^2+2\mathcal{E}_n}}{2\kappa T_{\frac{n-1}{2}}+\sqrt{(4T_{\frac{n-1}{2}}^2-1)\kappa^2+2\mathcal{E}_n}}\right] & n\text{ odd}\,,
\end{cases}
\label{eq:final_R}\\
&=
\begin{cases}
 -2\ln\kappa +\ln\left(\frac{2N+1}{N+1-n/2}\right) + \mathcal{O}(\kappa^2) & n\text{ even}\,,\\
 -\ln\kappa +\frac{n-1}{16(2N+1)} +\frac12 \ln\left(\frac{(7-4\sqrt{3})(2N+1)\left(2+\sqrt{4+\frac{1-n}{2N+1}}\right)^2}{n-1}\right) + \mathcal{O}(\kappa^2)& n\text{ odd}\,,
\end{cases}
\end{align}
with  $2\leq n \leq 2N+1$.
Keeping only the leading log terms, we find the compact expression
\begin{align}
R^\ast_{N,n} = R^\ast_{N,1} -
\begin{cases}
\left(\frac{3}{2}n -1\right)\ln \kappa+ \mathcal{O}(\kappa^0)  \,,& \text{ even } n\,,\\
\left(\frac{3}{2}n -\frac{3}{2}\right)\ln \kappa+ \mathcal{O}(\kappa^0)\,, &  \text{ odd } n\,,
\end{cases}
\end{align} 
for  $1\leq n \leq 2N+1$.
We also know $R^\ast_{N,1} = (2N+1)/\kappa^2$ for small $\kappa$, but a better prediction requires the subleading terms that we derive in the following section.

\subsection{Thick-Wall Limit}
\label{sec:thick}

The description of an excited-state profile in terms of transition functions determined by $2N+1$ radii satisfying $f''(R^\ast)=0$ is useful for small $\kappa$ but breaks down at larger $\kappa$.
Not only the shape of $f$ starts to differ, there also exists a threshold $\kappa$ beyond which $f''$ only has $N+1$ zeros; all $N$ radii with an even index in the above notation, $R^\ast_{N,2j}$, cease to satisfy $f''=0$, rendering our product of transition functions ansatz~\eqref{e.Nansatz} unsuitable. 
Rather than changing our radius definition, we content ourselves with predicting the odd radii for $\kappa\sim 1$. Since this covers the first and last radius, it provides a useful idea of the soliton's extent.

Near $\kappa= 1$, in the thick-wall limit, we use a different approximation from above. In this case, we define $\varepsilon=1-\kappa^2$ and consider $\varepsilon\ll1$. 
The potential can then be written as
\beq
V(f)=-\frac{\varepsilon}{2}f^2+f^4-\frac12f^6
\eeq
and therefore,
\beq
f''+\frac{2}{\rho}f'-\varepsilon f+4f^3-3f^5=0 \,.
\eeq
We then transform the equation into a rescaled profile and use a rescaled coordinate~\cite{PaccettiCorreia:2001wtt}:
\bea
f^2=\varepsilon g^2, \ \ \ \ \rho^2=z^2/\varepsilon\,.
\eea
The resulting equation is
\bea
g''+\frac{2}{z}g'-g+4g^3-3\varepsilon g^5=0\,,\label{e.gEq}
\eea
and of course the boundary conditions of $g$ are identical to $f$, that is $g'(0)=0$ and $g(\infty)=0$.

The leading order equation, with $\varepsilon\to0$, can be solved generally for any $N$. A simple technique to use is the shooting method. By choosing $g(0)$ with $g'(0)=0$ one can find approximate solutions to good accuracy. In Table~\ref{t.gsol}, we record the initial values $g(0)$ as well as the radii\textemdash defined by $f''=0$\textemdash for several $N$. The radii diverge for $\kappa\to 1$ according to
\begin{align}
R^\ast_{N,n} = \frac{c_{N,n}}{\sqrt{1-\kappa^2}}\,,
\label{eq:thickR}
\end{align}
with coefficients $c_{N,n}$ given in Tab.~\ref{t.gsol}.
Note that the coefficient of the $1/\sqrt{1-\kappa^2}$ term for the $N=0$ solitons matches very closely with the approximate fit found in Ref.~\cite{Heeck:2021zvk}.

\begin{table}[t]
\begin{tabular}[t]{c|c|c|c|c|c|c}
 \toprule 
 $N$ & $g(0)$ & $ c_{N,1}$ & $c_{N,3}$ & $c_{N,5}$ & $c_{N,7}$ & $c_{N,9}$  \\
\midrule\hline
0 & 2.168693539 & 0.345758 & $-$ & $-$ & $-$ & $-$ \\
\midrule\hline
1 & 7.051791599 & 0.106101 & 1.70188 & $-$ & $-$ & $-$  \\
\midrule\hline
2 & 14.565602713 & 0.0513571 & 0.793518 & 2.93172 & $-$ & $-$  \\
\midrule\hline
3 & 24.6803496815 & 0.0303081 & 0.464925 & 1.64239 & 4.15909 & $-$  \\
\midrule\hline
4 & 37.38615404998 & 0.0200077 & 0.306252 & 1.0686 & 2.58726 & 5.39492 \\
\bottomrule
 \end{tabular}
\caption{Initial value $g(0)$ for use in shooting-method solutions to Eq.~\eqref{e.gEq} with $\varepsilon=0$ as well as coefficient values for Eq.~\eqref{eq:thickR}. 
 \label{t.gsol} }
\end{table}

Table~\ref{t.gsol} also shows that a given odd radius $R_{N,2j+1}^\ast$ \emph{decreases} with $N$ for fixed $\kappa$, unlike in the small $\kappa$ limit where the radii \emph{increase} linearly with $N$. The reason is that for $\kappa\sim 1$ and small $N$, the particle does not start anywhere near the maximum of the potential, so friction can not be neglected and complicates the intuition. However, at large enough $N$, the particles are unavoidably pushed towards the maximum and our arguments from before apply; at large $N$, the radii again grow with $N$, although that is not captured by Eq.~\eqref{e.gEq} with $\varepsilon = 0$.

\subsection{Final Predictions for the Radii}

\begin{figure}[t]
\includegraphics[width=0.5\textwidth]{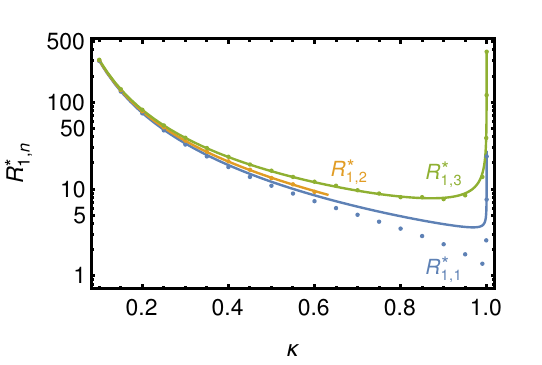}\hspace{-0.25cm}
\includegraphics[width=0.5\textwidth]{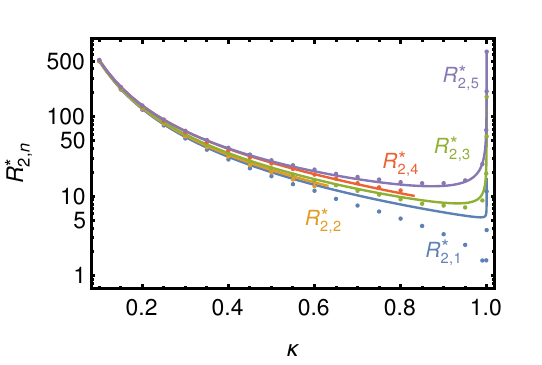}
\caption{
Left: the three radii $R^\ast_{1,n}$ vs.~$\kappa$ for the first excited state, $N=1$. Dots are numerical results and the solid lines the prediction from Eq.~\eqref{eq:final_radii}.
Right: radii for the $N=2$ excited state together with the prediction from Eq.~\eqref{eq:final_radii}.
 }
\label{fig:N1R}
\end{figure}

Having obtained approximations of the excited-state radii in the limiting cases of small and large $\kappa$ we can combine them to obtain approximate radius predictions valid for all $\kappa$:
\begin{align}
R^\ast_{N,n} = \frac{2N+1}{\kappa^2} +
\begin{cases}
 \frac{c_{N,n-1}}{\sqrt{1-\kappa^2}}-\left(\frac{3}{2}n -1\right)\ln \kappa  \,,& \text{ even } n\,,\\
 \frac{c_{N,n}}{\sqrt{1-\kappa^2}}-\left(\frac{3}{2}n -\frac{3}{2}\right)\ln \kappa \,, &  \text{ odd } n\,.
\end{cases}
\label{eq:final_radii}
\end{align} 
where the coefficients $c_{N,n}$ can be read off from Tab.~\ref{t.gsol} and $n=1,2,3,\ldots,2N+1$.
We mention again that the even radii, $R^\ast_{N,2k}$, no longer solve $f''(R^\ast_{N,2k})=0$ for large $\kappa$, so are no longer radii by our definition.
All radii for the first and second excited states are shown in Fig.~\ref{fig:N1R} and compared to numerical data. As expected, our predictions work well in the small and large $\kappa$ regimes in which they were derived, but differ somewhat for $\kappa$ in between.

\begin{figure}[t]
\includegraphics[width=0.5\textwidth]{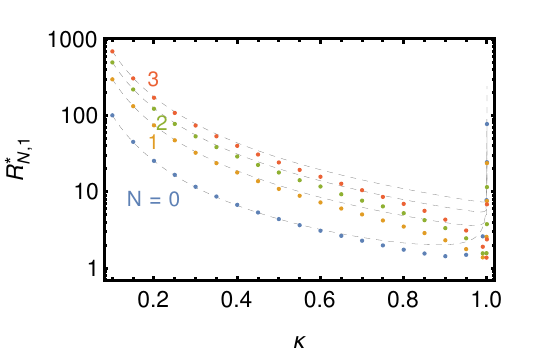}\hspace{-0.25cm}
\includegraphics[width=0.5\textwidth]{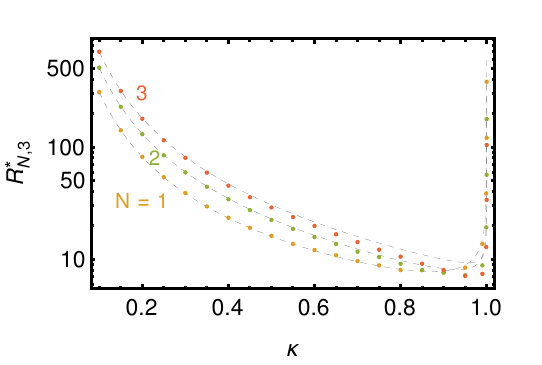}
\caption{
Transition radii $R_{N,1}^\ast$ (left) and $R_{N,3}^\ast$ (right)  vs.~$\kappa$ for several $N$. Gray dashed lines are the approximations from Eq.~\eqref{eq:final_radii}.}
\label{fig:radii}
\end{figure}

The general behavior of $R^\ast$ vs.~$N$ is shown in Fig.~\ref{fig:radii}. Notice that the naively expected behavior $R_{N+1,n} > R_{N,n}$ breaks down near $\kappa\sim 1$ because none of the particles are starting near the maximum $f_+$.  For large enough $N$, this behavior would be restored though.

\section{Charge, Energy, and Stability}
\label{sec:discussion}

For small $\kappa$, we are able to predict the profile $f(\rho)$ for any excited state $N$ to astonishing accuracy by combining the transition-profile ansatz from Eq.~\eqref{e.Nansatz} with our radius predictions from Eq.~\eqref{eq:final_radii}.
This then allows us to numerically calculate the integrals relevant for the Q-ball charge and energy, given in Eq.~\eqref{eq:QandE}. Since analytical approximations quickly become bothersome, we restrict ourselves to the leading terms in small $\kappa$ here.

Consider finding the value of the $Q$ integral. As shown in Eq.~\eqref{eq:QandE}, the nontrivial part takes the form
$
\int \dd\rho \,\rho^2 f^2
$
and in the $N=0$ case~\cite{Heeck:2020bau}, this takes the simple approximate value  of
$
\int \dd\rho \,\rho^2 f^2=R^{\ast3}/3
$
to leading order in large $R^\ast$.
If all the effects for an excited state $N$ with $2N+1$ transitions are included, we again find the leading-order result of
\beq
\int \dd\rho \,\rho^2 f^2=\frac{R^{\ast3}}{3}\,.
\label{eq:int_f_sq}
\eeq
 In this calculation we could have included a sum of $N$ discrete $\Delta\left(R^\ast\right)^2$ contributions where $\Delta$ is the width of the excited state shell around the center soliton, but as $\Delta\ll R^\ast$, these contributions are subleading.

For the surface-energy integral
$
\int \dd\rho \rho^2f^{\prime2}
$, 
 the leading-order behavior is different.
The $N=0$ result~\cite{Heeck:2020bau} is
$
\int \dd\rho \rho^2f^{\prime2}=R^{\ast2}/4 
$
to leading order in $R^\ast$.
For excited states, the effects of additional transitions need to be summed over, using our knowledge from above that the differences between the radii, $\Delta$, are large enough to separate the individual transitions, which yields
\beq
\int \dd\rho \rho^2f^{\prime2}=\left(2N+1 \right)\frac{R^{\ast2}}{4} \,.
\label{eq:int_fp_sq}
\eeq
It is significant that in this integral the $f^{\prime2}$ integral is proportional to the surface area at that radius, rather than a shell of volume.

\begin{figure}[t]
\includegraphics[width=0.5\textwidth]{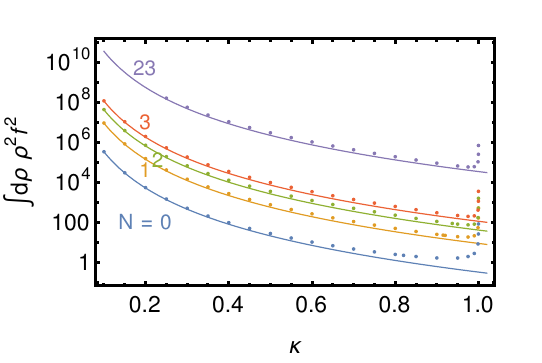}\hspace{-0.25cm}
\includegraphics[width=0.5\textwidth]{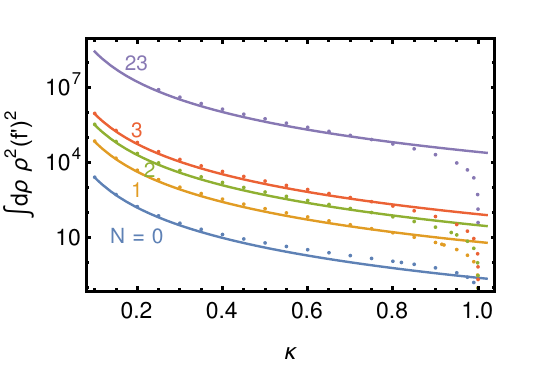}
\caption{
Integrals $\int \dd \rho \, \rho^2 f^2$ (left) and $\int \dd \rho \, \rho^2 f^{\prime 2}$ (right) vs.~$\kappa$\textemdash as relevant for Q-ball energy and charge\textemdash for the $N=0$ ground state and several excited states. The solid lines show the small-$\kappa$ approximations from Eq.~\eqref{eq:integrals}.
}
\label{fig:integrals}
\end{figure}

Overall, using the leading order relation $R^\ast=(2N+1)/\kappa^2$, we find
\begin{align}
\begin{split}
\int \dd\rho\,\rho^2f^2 & \simeq  \frac{(2N+1)^3}{3\kappa^6}\,,\\
\int \dd\rho\,\rho^2f^{\prime2} & \simeq \frac{(2N+1)^3}{4\kappa^4}\,,
\end{split}
\label{eq:integrals}
\end{align}
for small $\kappa$, which can also be obtained by inserting our radius predictions~\eqref{eq:final_radii} into the transition-profile ansatz~\eqref{e.Nansatz} and evaluating the integrals. For $N=0$, subleading terms have been derived in Ref.~\cite{Heeck:2020bau}.
The comparison with numerical data in Fig.~\ref{fig:integrals} shows that Eq.~\eqref{eq:integrals} is a fairly good approximation even for larger $\kappa$, and captures the $N$ dependence very well.
In fact, we find numerically that the scaling of the integrals with $(2N+1)^3$ is accurate far beyond the small-$\kappa$ regime: simply rescaling the $N=1$ integrals by $(2N+1)^3/(2+1)^3$ matches the $N=2\ldots 23$ integrals in the range $0<\kappa \lesssim 0.9$ to better than $10\%$ ($20\%$) for $\int \dd\rho\,\rho^2f^2$ ($\int \dd\rho\,\rho^2f^{\prime2}$).

The approximations in Eq.~\eqref{eq:integrals} as well as the general scaling with $(2N+1)^3$ break down near $\kappa\sim 1$, but we can still understand the numerical behavior qualitatively.
As shown above, in the thick-wall regime we expect the Q-ball radii to diverge according to $R^\ast\propto 1/\sqrt{1-\kappa^2}$. Since the integral $\int \dd\rho\,\rho^2f^2$ can be interpreted as the volume energy of the Q-ball~\cite{Heeck:2020bau}, we also expect this integral to diverge for $\kappa\to 1$. The \emph{surface} integral  $\int \dd\rho\,\rho^2f^{\prime2}$, on the other hand, decreases because the Q-ball profile becomes increasingly dilute for $\kappa\to 1$, suppressing the derivatives~\cite{Heeck:2020bau}.

\begin{figure}[t]
\includegraphics[width=0.49\textwidth]{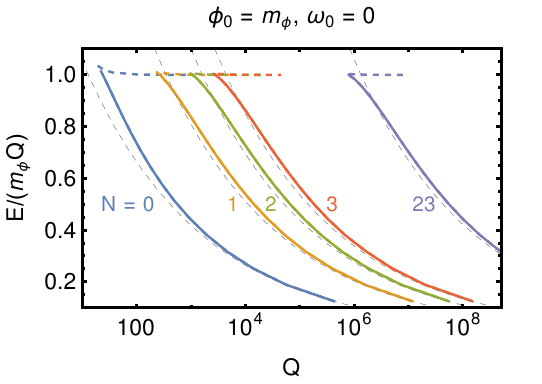}
\includegraphics[width=0.5\textwidth]{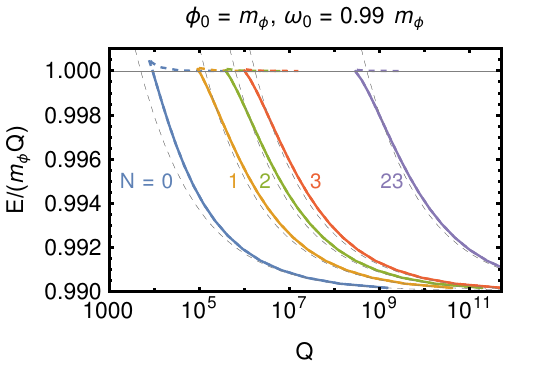}\\
\includegraphics[width=0.49\textwidth]{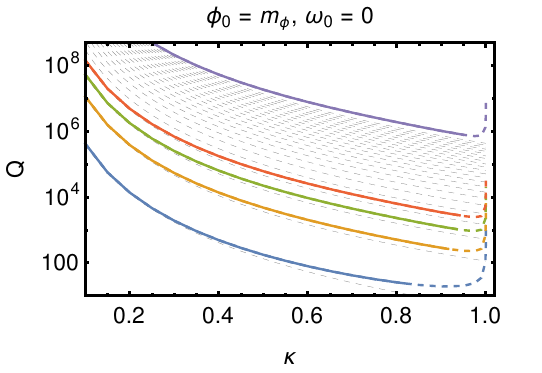}
\includegraphics[width=0.5\textwidth]{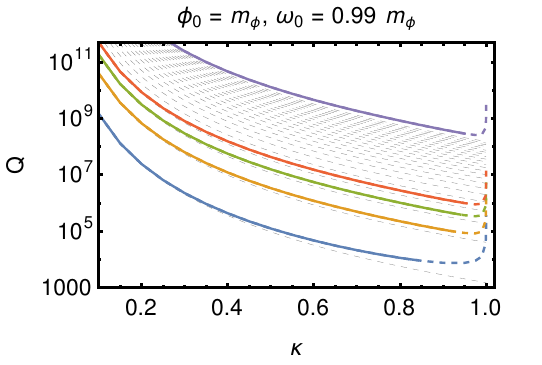}
\caption{
Top: $E/(m_\phi Q)$ vs.~$Q$ for $\omega_0=0$ (left) and $\omega_0 = 0.99 m_\phi$ (right) for the ground state ($N=0$) and several excited states. In the regime with $E/(m_\phi Q) > 1$, the Q-ball can decay into $Q$ free scalars, indicated by the dashed portion.
The gray dashed lines denote our analytical large-Q predictions given in Eq.~\eqref{eq:EvsQ} and footnote~2. 
Bottom: $Q$ vs.~$\kappa$ for the same examples. The gray dashed lines show the small-$\kappa$ prediction $Q\propto (2 N + 1)^3/\kappa^6$ prediction from Eq.~\eqref{eq:integrals} for $N=0$ to $N=23$ to illustrate the density of states. Note that we arbitrarily stop at $N=23$.
}
\label{fig:Qplots}
\end{figure}

Using the numerical integrals as a function of $\kappa$ or the small-$\kappa$ approximations, we obtain the Q-ball charge and energy via Eq.~\eqref{eq:QandE}. The numerical results are illustrated in Fig.~\ref{fig:Qplots} for some example parameters.
For small $\kappa$ and $\omega_0\neq 0$,\footnote{For $\omega_0=0$, the small-$\kappa$ approximation is quite different and to leading order in large $Q$ reads $E(Q) \simeq \frac52 \left(\frac{\pi}{3}\right)^{1/5}\left(N+\frac12\right)^{3/5}m_\phi^{3/5} \phi_0^{2/5} Q^{4/5}$.} we find an expression for $E(Q)$ that is valid for large $Q$:
\begin{align}
E(Q) \simeq \omega_0 Q + (2N+1) \left(\frac{\pi}{2}\right)^{1/3}\frac{3^{2/3} \sqrt{m_\phi^2 -\omega_0^2}}{2(\omega_0/\phi_0)^{2/3}} Q^{2/3}
\label{eq:EvsQ}
\end{align}
and shows that excited states simply have a $2N+1$ times larger surface energy than the ground state. 
Since the energy increases with $N$ for fixed $Q$, we are correct in denoting these states as excited states, and assume they eventually  decay into smaller Q-balls, as discussed below.\footnote{The charge-swapped Q-ball states explored in Refs.~\cite{Copeland:2014qra,Xie:2021glp} may be quasi-stable.}

In addition to energy, other components of the Q-ball energy--momentum tensor can also be approximated in the thin-wall limit. The scaling of several quantities with $N$ for a fixed $\kappa$ has been obtained in Ref.~\cite{Mai:2012cx}.

\subsection{ Stability of Excited States}

A detailed study of excited-state decays is postponed for future work, but let us nevertheless comment on stability. 
As shown in Ref.~\cite{Heeck:2020bau}, the $N=0$ states with $0<\kappa \leq \kappa_\text{stability}\simeq 0.8$ are stable because they satisfy $E < m_\phi Q$ (and $\dd Q/\dd \omega <0$~\cite{Tsumagari:2008bv}, which is a weaker stability criterion for our potential). The excited states with $N>0$ are, on the other hand, unstable~\cite{Friedberg:1976me} and can decay into smaller Q-balls and/or individual scalars.
From the above expression~\eqref{eq:EvsQ} it is clear that for sufficiently large $N$, the soliton will satisfy $E > m_\phi Q$, which then allows the Q-ball decay into $Q$ individual $\phi$ scalars. This instability occurs at large $\kappa$, with a threshold that increases with $N$ (see the dashed lines in Fig.~\ref{fig:Qplots}). Even more energetically favorable than a decay into individual scalars is a decay into ground-state Q-balls, for which many possible final states are kinematically allowed. At large $Q$, excited states with $\kappa\sim 1$ have $E\simeq m_\phi Q$, while the ground state has $E\simeq \omega_0 Q$ (for $\omega_0\neq 0$), so a configuration of $k$ ground state Q-balls with charges that add up to $Q$ still has a total energy around $\omega_0 Q$, which is smaller than $m_\phi Q$ by Eq.~\eqref{e.Omega0}. Phase space arguments generically prefer a decay into few states.

The, arguably more interesting and slightly more stable, Q-balls with $E < m_\phi Q$ are well described by our small-$\kappa$ expansion, at least for small $N$. These Q-balls cannot decay into $Q$ scalars, but produce Q-balls with smaller $Q$ in their final state; energetically, many final states are allowed~\cite{Mai:2012cx}. From Fig.~\ref{fig:Qplots} it is easy to see that for a given $N$ the condition $dQ/d\kappa<0$ holds for most $\kappa$. In analogy with the ground state analysis, this implies that these solitons cannot decay to other states of the same $N$. However, they can decay into states with smaller $N$, except for the stable $N=0$ case.
By again using Eq.~\eqref{eq:EvsQ} as an approximation, we can also estimate the number of those excited states for a fixed $Q$ as
\begin{align}
 N \simeq \frac{\sqrt{m_\phi^2 - \omega_0^2}}{3 m_\phi^{1/3} \phi_0^{2/3}} Q^{1/3} .
 \end{align}
While there are infinitely many excited states for a fixed $\kappa$, the physically more relevant restriction to fixed $Q$ only allows for a finite number of excitations. While angular excitations of Q-balls have been studied~\cite{Volkov:2002aj,Kleihaus:2005me}, to our knowledge the number of excitations for fixed $Q$ has not been specified. However, the known excited states satisfy $J=NQ$ for integers $N$ which suggests there are only a finite number of angular excitations with energies below a given value, such as $E<m_\phi Q$.

Let us focus on the $N=1$ excited state in the large-$Q$ regime with $\omega_0\neq0$, so that we may use Eq.~\eqref{eq:EvsQ} as an approximation.
One possible decay mode (and preferred by phase space considerations) is into two Q-balls of charge $Q_1$ and $Q-Q_1$, which have a lower energy of
\begin{align}
E(Q_1)_{N=0}+E(Q-Q_1)_{N=0} -E(Q)_{N=1} &\simeq  -\left(\frac{\pi}{2}\right)^{1/3}\frac{3^{2/3} \sqrt{m_\phi^2 -\omega_0^2}}{2(\omega_0/\phi_0)^{2/3}}\nonumber\\
&\quad \times\left(3Q^{2/3}-(Q-Q_1)^{2/3} - Q_1^{2/3}\right).
\end{align}
The energy gain is minimized for $Q_1 = Q/2$, i.e.~a decay into two ground-state Q-balls of equal charge, making this the least likely decay judging by phase space. More energy is released for $Q_1\ll Q$, where our thin-wall expression from Eq.~\eqref{eq:EvsQ} breaks down for the smaller Q-ball. Let us consider the extreme case of the $N=1$ decay into a $N=0$ Q-ball of charge $Q-1$ and a free scalar $\phi$, with energy gap
\begin{align}
E(Q-1)_{N=0}+E_\phi -E(Q)_{N=1} &\simeq  -2 Q^{2/3}\left(\frac{\pi}{2}\right)^{1/3}\frac{3^{2/3} \sqrt{m_\phi^2 -\omega_0^2}}{2(\omega_0/\phi_0)^{2/3}}\,,
\end{align}
ignoring kinetic energies, so $E_\phi \simeq m_\phi$. For large $Q$, the $\phi$ mass is negligible so the same expression holds for the emission of additional scalars or scalar--anti-scalar pairs. 
Phase space will once again prefer a decay into few particles.
Since the mass gap grows with $Q^{2/3}$, the lifetimes of large excited states are expected to be parametrically short, although the actual numbers depend on the values for the potential parameters.

\section{Generalization To Other Potentials\label{sec:GenPotential}}
So far our treatment of excited states has focused on the sextic potential of Eq.~\eqref{e.sextic} as a definite example. In this section, we outline what aspects of our analysis are expected to be robust over many potentials that admit a thin-wall limit. We also highlight where differences between potentials play a significant role. These generalizations assume a single scalar field, leaving the analysis of multi-field models to future work. 

We return, for the moment, to the basic requirements on the scalar potential for Q-ball solutions. It is useful to rewrite the conditions discussed in Sec.~\ref{sec:ground} on $U(|\phi|)$ in  terms of the dimensionless potential
\beq
\widehat{U}(f)=\frac{1}{\phi_0^2(m_\phi^2-\omega_0^2)}U(f\phi_0/\sqrt{2})~.
\eeq
We find that
\beq
\widehat{U}(0)=0~, \ \ \ \ \left.\frac{\dd\widehat{U}}{\dd f}\right|_{f=0}=0~, \ \ \ \ \left.\frac{\dd^2\widehat{U}}{\dd f^2}\right|_{f=0}=\frac{m_\phi^2}{m_\phi^2-\omega_0^2}~.
\eeq
The condition that $U(|\phi|)/|\phi|^2$ have a minimum at $\phi_0/\sqrt{2}$ is what allows for thin-wall solitons. This requirement is expressed as
\beq
\widehat{U}(1)=\frac{\omega_0^2}{2(m_\phi^2-\omega_0^2)}~, \ \ \ \ \left.\frac{\dd\widehat{U}}{\dd f}\right|_{f=1}=\frac{\omega_0^2}{m_\phi^2-\omega_0^2}~, \ \ \ \ \left.\frac{\dd^2\widehat{U}}{\dd f^2}\right|_{f=1}>\frac{\omega_0^2}{m_\phi^2-\omega_0^2}~.
\eeq
We can then determine the extrema of $V(f)$, the effective potential that appears in the equations of motion. The locations of the extrema are determined by
\beq
\frac{\dd\widehat{U}}{\dd f}=f\frac{\omega^2}{m_\phi^2-\omega_0^2}~.
\eeq
The second derivative of $V(f)$ reveals maxima and minima:
\beq
\frac{\dd^2V}{\dd f^2}=\frac{\omega^2}{m_\phi^2-\omega_0^2}-\frac{\dd^2\widehat{U}}{\dd f^2}~.
\eeq
For instance, we have an extremum  at $f=0$ with
\beq
\left.\frac{\dd^2V}{\dd f^2}\right|_{f=0}=\frac{\omega^2-m_\phi^2}{m_\phi^2-\omega_0^2}\leq0~.
\eeq
This is less than zero because $\omega\leq m_\phi$, so this is a maximum except at $\omega=m_\phi$. We also see that $V(f)$ has an extremum at $f=1$, but only when $\omega=\omega_0$, the thin-wall limit. In this case
\beq
\left.\frac{\dd^2V}{\dd f^2}\right|_{f=1}=\frac{\omega_0^2-m_\phi^2}{m_\phi^2-\omega_0^2}=-1~,
\eeq
so this is also a maximum. Clearly, there must be at least one minimum between these two maxima. However, depending of the specifics of $U(|\phi|)$ there may be other extrema as well. The salient point is that any potential that satisfies the requirements to support thin-wall Q-balls has a local maximum at $f=1$ when $\omega=\omega_0$.

By construction, $V(0)=0$. We also have $V(1)=0$ when $\omega=\omega_0$. This means that in the $\omega=\omega_0$ limit, the ``particle" may transition from the $f=1$ maximum and end at the $f=0$ maximum only after waiting an infinite amount of ``time" (related to infinite radius for the Q-ball) before transitioning. When it does transition it rolls without friction, meaning there is a conserved energy
\beq
\mathcal{E}=\frac12f^{\prime2}+V(f)~.
\eeq
 This can be extracted directly from the equations of motion~\eqref{eq:differential_equation} in the infinite $\rho$ limit
\beq
\frac{\dd}{\dd\rho}\left(\frac12f^{\prime2}+V(f)\right)=0~.
\eeq
In this ``thin-wall" limit the particle is at rest at $\rho=0$ ($f=0$) and at $\rho=\infty$ ($f=1$). Therefore, both of these points have $\mathcal{E}=0$. This special trajectory, what we have called the transition function $f_T$, satisfies
\beq
\frac{\dd f_T}{\dd\rho}=-\sqrt{-2V(f_T)}~.
\label{eq:general_transition}
\eeq
The negative root must be chosen to ensure $f_T$ decreases with increasing $\rho$.

For the sextic potential this equation leads to Eq.~\eqref{eq:fT}, but it is well defined for any potential that gives rise to Q-balls. 
Albeit generally impossible to solve analytically, Eq.~\eqref{eq:general_transition} is trivial to integrate numerically, providing an approximate thin-wall solution to the original differential equation. 
As shown above, these transition functions are the building blocks from which radial Q-ball modes are constructed. In this paper our numerical results show this explicitly with the sextic potential. However, our final approximate analytical result for the $N$-th excited state profile~\eqref{e.Nansatz} is given in terms of $f_T$ to indicate how our the results that follow from this profile are expected to generalize to other potentials. 

In the thin-wall limit, $\omega\to\omega_0$, the profile form given in \eqref{e.Nansatz} must be correct.
In this case, the particle rolls without friction and makes $N$ transitions without friction between the maxima at $f=\pm1$  and $f=0$. Each of these transitions is given by the function defined above. We have seen that for the sextic potential they remain a good approximation well away from this limit, but for more complicated potentials this agreement may break down sooner.

Despite being derived for our particular sextic potential, we expect the $E(Q)$ result of Eq.~\eqref{eq:EvsQ} to be approximately valid for all potentials that exhibit a thin-wall limit, since it relies mainly on the $N$ scaling of the integrals in Eqs.~\eqref{eq:int_f_sq} and~\eqref{eq:int_fp_sq}. For large $Q$ and small $N$, radially excited states simply increase the Q-ball energy by $2N$ times the surface energy. The spectrum is surprisingly simple and can be estimated using ground-state properties (volume and surface energies).

\section{Conclusion\label{s.con}}
Q-balls are simple non-topological solitons that can be interpreted as bound states of $U(1)$-symmetric scalars. The ground state solutions in a wide range of parameter space are stable because they furnish the smallest-energy configuration for a fixed $U(1)$ charge $Q$. As with other bound-state systems, excited Q-balls states with energies above the ground state can exist as well. 

These excited states can play an essential role in accurately describing soliton production as well as the dynamics of soliton scattering. From the standpoint of these physical motivations the excited state spectrum for fixed charge $Q$ is most important. We have provided the first characterization of this spectrum, focusing on the sextic potential. In particular, we have shown that Q-balls of fixed $Q$ have a finite number of radial excitations. This number grows with $Q$ and some smaller solitons have only a few excited states of this type or even none at all.

We have developed a qualitative understanding of the radial excitations of Q-ball solitons and provided accurate approximate solutions.
Our results are sufficient to describe the properties of arbitrary excited states in the sextic potential and should carry over with minimal modifications to other cases. While we have commented on the (in)stability of these excited states the difficult task of calculating the decay rates into smaller ground-state Q-balls and/or individual scalars is left for future work.
The analytic approximations obtained here for global Q-balls can immediately be used to describe excited gauged and Proca Q-balls via the mapping of Refs.~\cite{Heeck:2021zvk,Heeck:2021bce}, to be discussed in detail elsewhere. 

\section*{Acknowledgements}
This work was supported in part by NSF Grant No.~PHY-1915005. The research of Y.A.~was supported
by Kuwait University.

\bibliographystyle{utcaps_mod}
\bibliography{BIB}

\end{document}